\documentclass[aps, pra, preprint, groupedaddress, amsfonts,
               amsmath, amssymb, showpacs, nofootinbib]{revtex4-1}
\usepackage{microtype}
\usepackage{graphicx}
\usepackage{epstopdf}
\usepackage[utf8]{inputenc}
\usepackage[T1]{fontenc}
\usepackage[usenames,dvipsnames]{xcolor}
\usepackage{hyperref}
\usepackage{booktabs}

\usepackage{color}

\begin{document}
\title{A generalized independent atom model approach for net ionization of molecules by multiply-charged heavy-ion impact}

\author{Hans J\"urgen L\"udde}
\email[]{luedde@itp.uni-frankfurt.de}
\affiliation{Center for Scientific Computing, Goethe-Universit\"at, D-60438 Frankfurt, Germany} 

\author{Marko Horbatsch}
\email[]{marko@yorku.ca}
\affiliation{Department of Physics and Astronomy, York University, Toronto, Ontario M3J 1P3, Canada}

\author{Tom Kirchner}  
\email[]{tomk@yorku.ca}
\affiliation{Department of Physics and Astronomy, York University, Toronto, Ontario M3J 1P3, Canada}
\date{\today}
\begin{abstract}
The previously applied independent atom model (IAM) for highly charged ion-molecule collisions which implemented the suppression of 
multiple ionization and capture on the basis of geometric overlaps of cross-sectional areas representing ion-atom cross sections
using a pixel counting method (PCM), [Phys. Rev. A {\bf 101}, 062709 (2020)] is extended to incorporate the possibility of multiple collisions within the molecule.
This is accomplished on the basis of estimated mean free paths for sequential projectile-atom collisions.
The IAM-PCM was demonstrated to be successful in describing proton-molecule collisions, and moderately-charged ion impact
at high collision energies. The new model does agree with these results, but has important consequences for highly charged projectiles
providing larger cross sections than IAM-PCM, but still well below the simple additivity rule results.
\end{abstract}
%
%

\maketitle
\section{Introduction}
\label{intro}
Multiple ionization and capture of electrons by ions colliding with complicated molecules (e.g., biomolecules, or molecules of interest in astrophysics) which are
made up of atomic hydrogen (H) and second-row atoms (C, N, O, F), and other atoms represents an important problem even at the level of net cross sections.
These collisions, particularly at medium to high collision energies, i.e., energies where the projectile velocity is in the range of target electron velocities can be 
described by a semiclassical approach where the motion of the nuclei is treated classically, while the electron motion is described by quantum mechanics.

The description of such processes focuses often on cross sections differential in electron emission angle and energy to be compared with experimental
data. Among the methods describing the electron motion quantum mechanically are those based on the Born approximation, e.g., the work of 
Ref.~\cite{Cappello_2008}, which also uses the complete neglect of differential overlap (CNDO) method
to represent molecular orbitals in terms of atomic orbitals. This allows the molecular cross sections to be decomposed in terms of atomic
orbital contributions.
For total electron production cross sections at intermediate and low energies one needs a non-perturbative approach, however.
One class of approaches where the electron motion is described using classical statistical mechanics in the sense of an $\hbar=0$ approximation
to the quantum Liouville equation is the classical-trajectory Monte-Carlo method (CTMC). Biomolecules were described by this methodology,
e.g., in Refs.~\cite{PhysRevA.92.062704, Sarkadi_2016}, and for simpler molecules, e.g., $\rm H_2O$ in Ref.~\cite{PhysRevA.99.062701}.
These were carried out in an independent-electron model (IEM) approximation. At the level of explicit many-electron treatment an $N-$ particle
CTMC model was applied to deal with multiple ionization~\cite{PhysRevA.106.012808}.

An improved quantum treatment of the electronic motion is obtained within the continuum distorted wave approximation
with eikonal initial state (CDW-EIS) which predicts both differential and total cross sections much better than the Born
approximation~\cite{Itoh13}. 
Recent developments involve some modelling of the target charge for the outgoing wave,
and  comparison has been made with experiments and IEM-CTMC results  for the water molecule at intermediate energy~\cite{PhysRevA.105.062822}.
A more detailed description can be found in Ref.~\cite{PhysRevA.111.042815}. The CDW-EIS method borrows from quantum chemistry
the molecular orbitals of the molecule, usually via the CNDO coefficients. More sophisticated CDW-EIS implementations have been reported
in Refs.~\cite{PhysRevA.93.032704} and~\cite{PhysRevA.108.032815}.

Independent atom models (IAM) have been applied in this context in various forms. The simplest form is the additivity rule (AR) where the cross sections
for projectile collisions with constituent atoms are simply added together. The IAM-AR approach relies on previously calculated ion-atom collisions. It usually
overestimates ion-molecule ionization cross sections  in the vicinity of the Bragg peak, but is found to be valid at extremely high collision energies. 
In order to address this problem the IAM pixel counting method (PCM) was introduced in Ref.~\cite{hjl16} and applied to net ionization and net capture.
The basic philosophy is to make use of highly accurate non-perturbative projectile-atom cross sections using the so-called basis generator method (BGM).
A comparison of such atomic cross sections with other state-of-the-art results can be found, e.g., in Fig.~1 of Ref.~\cite{hjl20} for atomic hydrogen targets,
and for other target atoms of interest in Ref.~\cite{hjl25a,hjl25b}. Many-electron target atoms are described at the level of the
optimized potential method of density functional theory which is a local effective potential to approximate Hartree-Fock theory.
It does include electronic exchange, but not correlation. 

The IAM-PCM approach introduced geometric screening corrections  by considering trajectories of projectiles colliding with molecules with different orientations. For each orientation the atomic constituents were represented by disks with areas proportional to the given net cross section for
the atom. The overlapping areas were determined for each orientation 
by a pixel counting method, and then used to determine the ion-molecule cross sections by averaging over orientations. 
Applications were made for proton collisions with few-atom molecules,  with  biomolecules and with some clusters in Refs.~\cite{hjl18, hjl19, hjl19b, hjl20, hjl22}.  Recently, PCM has been applied in Ref.~\cite{Bernal2024} to electron capture by protons from biomolecules on the basis
of proton-atom collision cross sections obtained with a numerical solver for time-dependent density functional theory. While some discrepancies with our PCM results were reported it turns out that these were caused by a programming error.

The field of electron or positron collisions from molecules has also experienced applications of IAM approaches, in particular the so-called screening corrected
SCAR method described in Refs.~\cite{Garcia_2010, Garcia_2016}. Recently the method has been used to describe ionization in positron-molecule
collisions with a correlated description of atomic fluorine~\cite{Mori_2024}.

In Ref.~\cite{hjl20a} the IAM-PCM was applied to collisions of charged ions for a wide range of projectile charges with gas-phase water and  
uracil ($\rm C_4 H_4 N_2 O_2$)
for which experimental and other theoretical results were available. High-quality ion-atom cross sections from the two-center basis generator method 
(TC-BGM) for net ionization 
were presented for projectile charges $Q=1,2,3$ and a scaling method was established to deal with higher projectile charges. 
For the uracil target experimental data were explained for charge range $Q=1..8$ and for water vapor up to $Q=13$. 

The objective of the present work is to introduce a multiple-scattering approach to projectiles colliding with atoms within a molecule.
As is true for other strict IAM approaches, the only input from quantum chemistry is the geometry of the molecule.
This is obtained by generalizing PCM to include double-scattering events on the basis of a mean free path model.
To the best of our knowledge this is the first time that multiple scattering is implemented in this context.


The paper is organized as follows. In Sect.~\ref{sec:model} we introduce the theoretical basis for the current work. Sect.~\ref{subsec:meanfreepaths}
describes the determination of mean free paths for projectile collisions with constituent atoms within the molecule and provides some examples.
Sect.~\ref{subsec:pixel} provides the new PCM methodology (xPCM) for assigning pixels as a function of atom-atom distance to account for multiple scattering.
Since the work focuses solely on an IAM treatment of collisions, 
we will drop the IAM acronym, i.e., PCM and xPCM stand for IAM-PCM and
IAM-xPCM respectively.
In Sect.~\ref{sec:expt} we compare the xPCM results with PCM and with experiment: Proton collisions are presented in Sect.~\ref{subsec:expt1}, collisions
with charged ions in Sect.~\ref{subsec:expt2} for water molecules, adenine and uracil, and anthracene.
The paper ends with a few concluding remarks in Sect.~\ref{sec:conclusions}.

\section{Model}
\label{sec:model}
The technique of the PCM implements the overlap of circular areas which represent the atomic cross section as a sequence along
the projectile path for a given molecular orientation by assigning one pixel without consideration of  the distance between the
circles. The implication is that only a single ionization (or capture) event occurs along the chosen path.
Since it is not recorded which of the atoms caused the event, all of the pixels representing the encountered atoms are weighted equally.
For the case of $m$ overlapping atomic cross sections each participating atom is assigned a pixel fragment of size $m^{-1}$.
The reduction of the AR molecular cross sections to the PCM results can be interpreted as the transfer of the single-collision condition
of the projectile-molecule system to the level of projectile collisions with constituent atoms.

The proposed generalization of PCM intends to allow for multiple collisions within the molecule. Thus, the molecular collision is
described as a sum of independent projectile-atom collisions. In order to implement multiple scattering conditions properly we
need to take into account for the following fact: the number density of atomic constituents within the molecule may be so large that
the mean free path between projectile-atom collisions is smaller than the relevant interatomic distance along the chosen projectile path.
Therefore, multiple projectile-atom scattering conditions within the molecule may occur. 

\subsection{Determination of mean free paths}
\label{subsec:meanfreepaths}

Let us define a mean free path for a given projectile with collision energy $E$ in a given atomic medium
\begin{equation}
\lambda(E) = \frac{1}{n\;\sigma(E)} \ .
 \label{eqn:eq1} 
\end{equation}
Here $\lambda(E)$ is given in units of $\mbox{\normalfont\AA}$, $n$ is the number density of the given constituent atom 
of the molecule,
and $\sigma(E)$ is the atomic collision cross section for the process of interest which is net ionization for the present work.
To provide an example in the context of the IAM description: $n$ could represent the number density of carbon atoms 
within the molecular volume (in $\mbox{\normalfont\AA}^{-3}$),
and $\sigma(E)$ the net ionization (or capture) cross section for projectile-carbon collisions (in $\mbox{\normalfont\AA}^{2}$).

The first task is to find the number density for a given atom within the molecular volume. We define the van der Waals (vdW) molecular volume
by considering the packing of the constituent atomic vdW spheres. The calculation is performed by discretization of the volume: one begins
by defining a clipping volume that contains the molecule; this volume is discretized by small cuboids (voxels); for each voxel one determines
whether it is empty, or whether at least one of the vdW atomic spheres has some occupancy within the voxel. The molecular vdW volume is obtained
from the sum of all occupied voxels with geometric center within the molecule. The calculation is carried out at a few levels of resolution
to determine the accuracy of the calculated volume.

\begin{table}[h!]
 \centering
 \begin{tabular}{cccccccc}
 \toprule
 {\bf molecule} & {\bf formula} &{\bf vdW vol.}&{\bf $n_{\bf C}$} & { $n_{\bf H}$} & { $n_{\bf N}$} & { $n_{\bf O}$}& {\bf group} \\
 \midrule
   adenine & C$_5$H$_5$N$_5$ & 107.1 &  0.047 &  0.047 &  0.047 & - & purine \\
   uracil & C$_4$H$_4$N$_2$O$_2$ & 88.01 &  0.045 &  0.045 &  0.023 & 0.023 & pyrimidine\\ 
   tetrahydrofuran& C$_4$H$_8$O & 71.50 &  0.056 &  0.112 &  - & 0.014 & cyclic ether\\ 
   valine & C$_5$H$_{11}$NO$_2$ & 108.7 &  0.046 &  0.101 & 0.009 & 0.018 & amino acid\\ 
   anthracene & C$_{14}$H$_{10}$ & 163.7 &  0.085&  0.061 &  - & - & PAH\\               
    \bottomrule
    \end{tabular}
 \caption{Atomic number densities for carbon, hydrogen, nitrogen and oxygen (in $\mbox{\normalfont\AA}^{-3}$) for some molecules. The molecular vdW volumes (in $\mbox{\normalfont\AA}^{3}$) 
 were calculated numerically to four-digit accuracy as explained in the text. }
 \label{tab:Teilchenzahldichten}
\end{table}

\begin{figure}[h!]
  \centering
  \begin{minipage}[t]{0.5\textwidth}
   \centering
   \includegraphics[width=\textwidth]{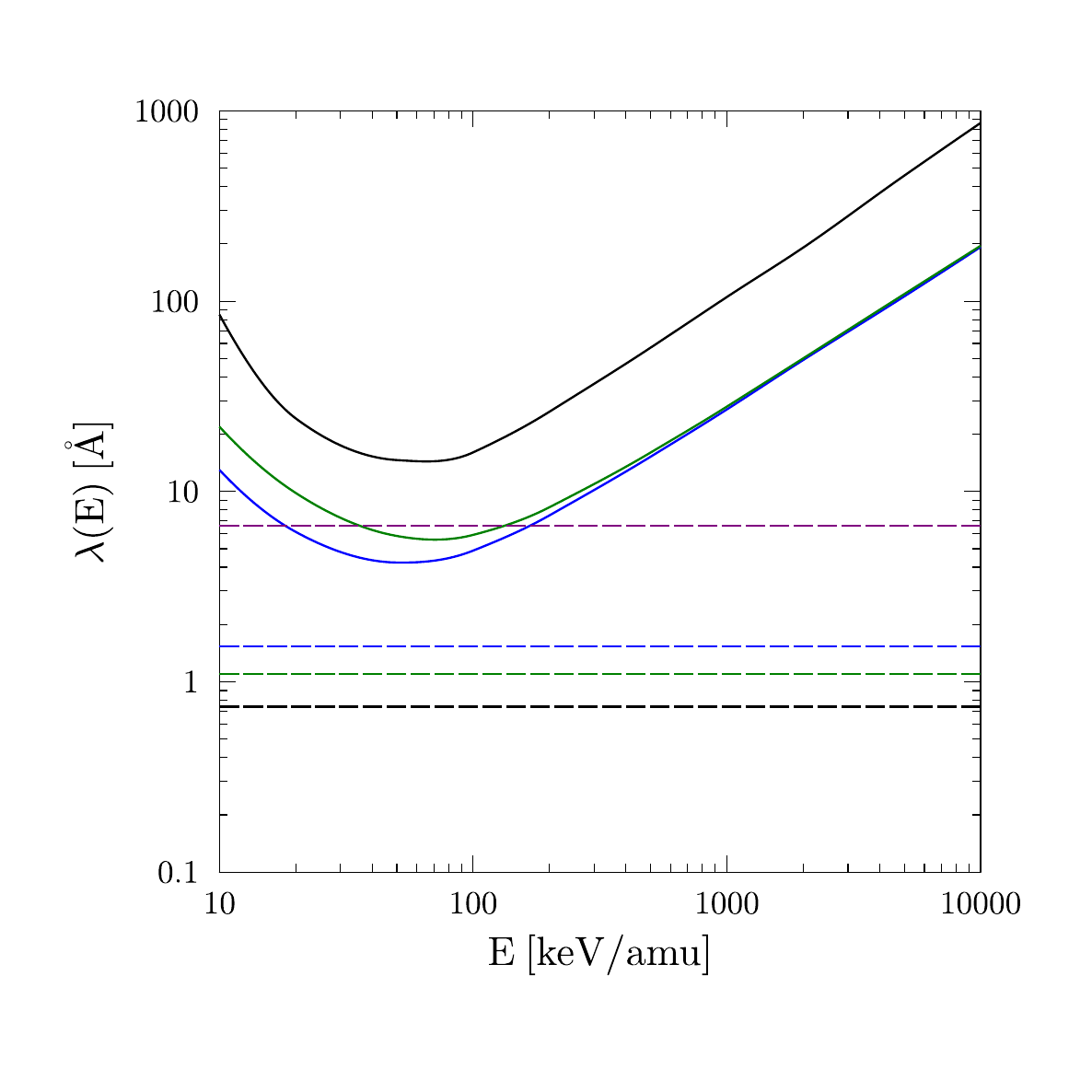}
  \end{minipage}
 \hskip -1 true cm 
    \begin{minipage}[t]{0.5\textwidth}
   \centering
   \includegraphics[width=\textwidth]{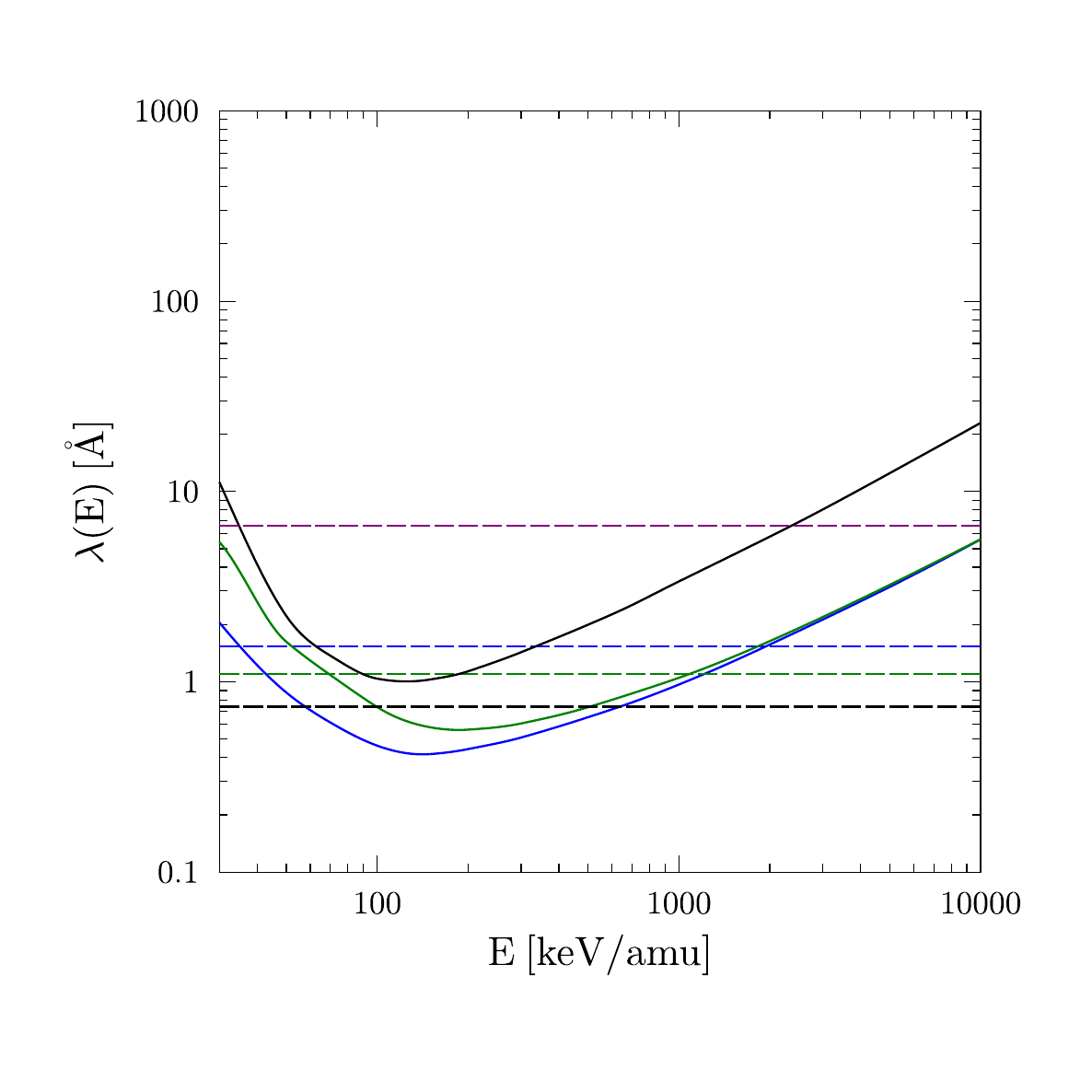}
  \end{minipage}
  \hfill  
   \caption{Mean free paths [Eq.~(\ref{eqn:eq1})] for ionizing collisions with adenine: in the left panel for proton projectiles, and in the right panel for bare carbon ion projectiles.
   The solid curves show $\lambda(E)$ for H (black), C (blue), N (green), while the corresponding dashed lines are the
   bond lengths for diatomic molecules. The purple dashed line shows the largest interatomic distance length for the molecule.} 
    \label{fig:lambdaAdenin}    
\end{figure} 

Table~\ref{tab:Teilchenzahldichten} lists the calculated vdW volumes and number densities for a few typical molecules which belong
to different groups. 
Note that the constituent atoms are treated as independent (hence IAM), and thus each atomic species (H, C, N, O)
has a number density which is specific for each chosen molecule.
These molecules have typically linear dimensions of less than 10~\AA . 
atoms are primary candidates for ionizing multiple scattering.
We can expect only a minimal multiple-scattering effect for collisions of protons with these molecules, i.e., the previously
obtained PCM cross sections should not be affected. The situation changes for collisions of charged ions and multiple
scattering may become relevant for high projectile charges. The increase in net ionization cross sections leads to a reduced mean
free path for a substantial range of collision energies (e.g., in the Bragg peak region), and so they can become smaller than
typical atomic separations within the molecule. We should expect multiple scattering to gain importance in such cases.

We focus on net ionization in this work. In general, the mean free path depends on: {\it (i)} the given molecule,
{\it (ii)} the atomic species via its number density
 ; {\it (iii)} the process considered; {\it (iv)} the collision energy;
and {\it{(v)}} the projectile charge. If $\lambda(E)$ becomes less than the distance between two atoms with overlapping cross sectional
areas represented by disks, multiple scattering will occur and should be taken into account.

For proton collisions it follows that only carbon
atoms are primary candidates for ionizing multiple scattering.
We can expect only a minimal multiple-scattering effect for collisions of protons with these molecules, i.e., the previously
obtained PCM cross sections should not be affected. The situation changes for collisions of charged ions and multiple
scattering may become relevant for high projectile charges. The increase in net ionization cross sections leads to a reduced mean
free path for a substantial range of collision energies (e.g., in the Bragg peak region), and so they can become smaller than
typical atomic separations within the molecule. We should expect multiple scattering to gain importance in such cases.

\begin{figure}[h!]
  \centering
  \begin{minipage}[t]{0.5\textwidth}
   \centering
    \includegraphics[width=\textwidth]{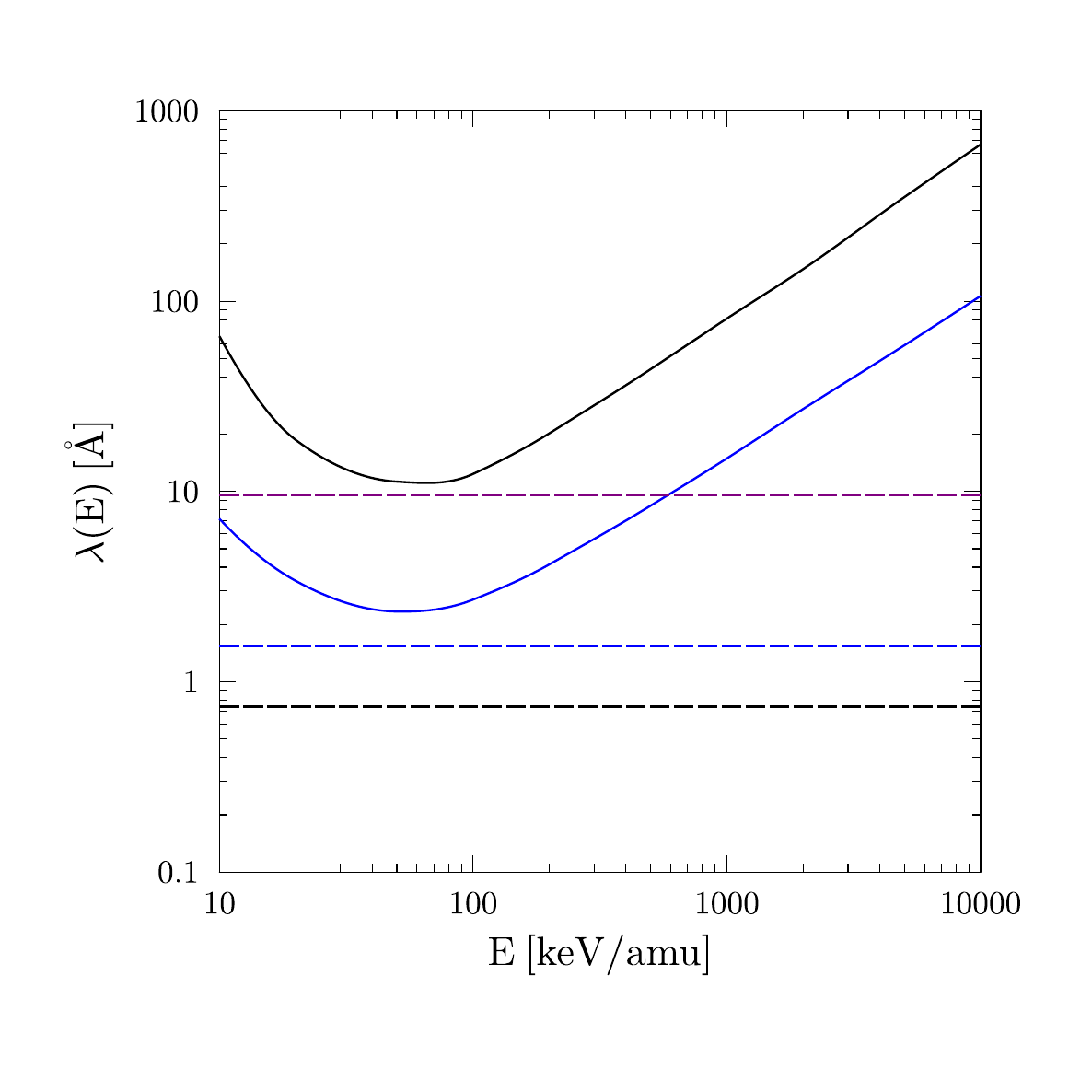}
  \end{minipage}
  \hskip -1 true cm 
   \begin{minipage}[t]{0.5\textwidth}
   \centering
   \includegraphics[width=\textwidth]{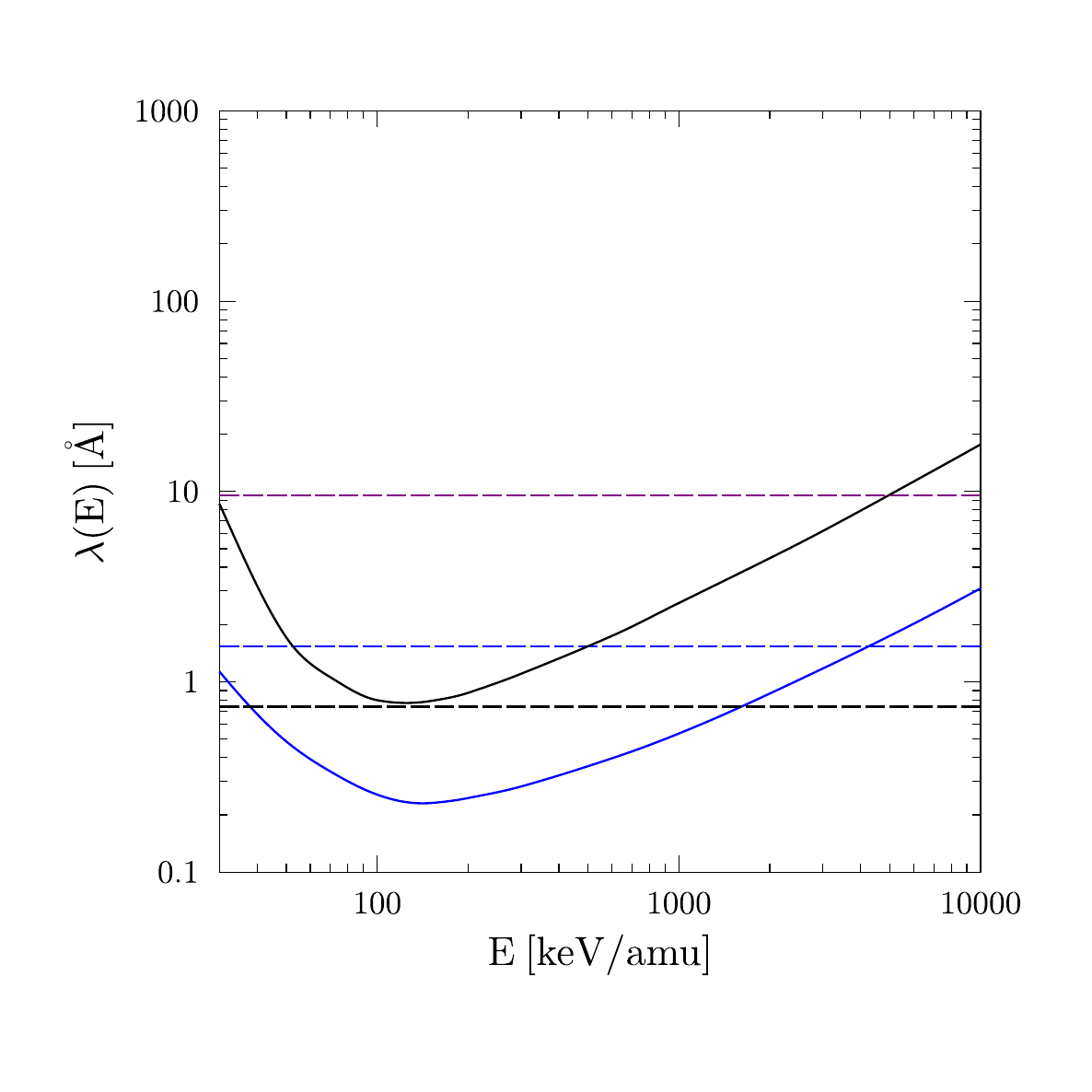}
  \end{minipage}
  \hfill       
   \caption{Same as in Fig.~\ref{fig:lambdaAdenin}, but for anthracene target molecules.}
   \label{fig:lambdaAnthracen} 
\end{figure}

Some examples of the dependence of the mean free path on collision energy and on projectile charge
are provided in Fig.~\ref{fig:lambdaAdenin} for adenine and Fig.~\ref{fig:lambdaAnthracen} for anthracene. 
For adenine we observe in the left panel of Fig.~\ref{fig:lambdaAdenin} that ionization in proton collisions is unlikely
to produce multiple ionizing scattering events, since the mean free paths for carbon and nitrogen constituents even at their minimum
do not fall significantly below the relevant interatomic distance scale (magenta dashed line at 6.62~\AA). 
This is different for bare carbon projectiles: the right panel of Fig.~\ref{fig:lambdaAdenin}
shows that there is a considerable energy range for which the mean free paths for carbon and nitrogen dip well below
not only the linear scale of the molecule (magenta dashed line) but even the typical diatomic bond lengths.
These bond lengths are only included for reference here to provide some understanding of the mean free path lengths.

Fig.~\ref{fig:lambdaAnthracen} shows the situation for anthracene. This molecule is different in the respect that the carbon atoms form the
center of a planar triple ring structure surrounded by hydrogen atoms. As a result there is a higher chance of multiple scattering
from C atoms if the trajectory falls within the molecular plane. We observe in the left panel of Fig.~\ref{fig:lambdaAnthracen} that even in 
the case of proton collisions the mean free path for ionizing carbon falls below the magenta dashed line at 9.53~\AA, which represents the extent of the molecule.
For bare carbon impact (right panel) even the collisions from different hydrogen atoms can lead to multiple ionization.

\subsection{Pixel distribution for multiple scattering}
\label{subsec:pixel}

We now propose a method of how to generalize PCM to include ionization from multiple scattering events.
Consider a molecular orientation for which $m$ atomic cross sectional disks overlap (at least partially).
In the plane perpendicular to the beam direction consider a point with coordinates $(x,y)$ and assume that there are
$m$ pixels corresponding to different atoms separated along the chosen trajectory along the $z$-direction. In the semiclassical approximation
we have a projectile trajectory which allows to distinguish the atoms as a sequence of events and we can enumerate
those along the $z$-axis which represents the projectile path.
For double-scattering events we consider pairs of pixels $(i,j)$ with $i<j$, and we denote the separation along the 
$z$-axis as $d_{i,j}$, and the relevant mean free path for the collision as $\lambda_i(E)$.
The mean free path is labeled by $i$, since it is chosen for the first encountered atom.

We illustrate the procedure with an example of scattering from an O-H bond within the molecule. If oxygen is encountered
first, we use $\lambda_{\rm O}$ which is smaller than $\lambda_{\rm H}$ 
(in analogy to the examples shown, e.g., in Fig.~\ref{fig:lambdaAdenin}). Thus, double-scattering can occur for this 
sequence of events, but not for the reverse process, where the hydrogen atom is encountered first, if 
$\lambda_{\rm H}$ is too large compared to $d_{i,j}$.

We assign the following rules for PCM and extended PCM (xPCM):

\vspace{-0.5cm}
\begin{itemize}
\item
pixel assignment for single collisions, i.e., local pixel assignment (PCM limit): for each atomic center assign $1/m$ pixels;
\item
pixel assignment for double collisions, i.e., non-local pixel assignment, i.e., xPCM: for each pair $(i,j)$ with $i<j$ and $d_{i,j}> \lambda_i$
assign $1/m$ pixels for the partner $j$ in addition to the PCM limit pixel assignment.
\end{itemize}

With the xPCM prescription the total pixel sum along an overlapping sequence is obtained as
\begin{equation}
  m_{\rm pix} = {1} +{ \frac{1}{m}\sum_{i=1}^{m-1}\sum_{j=i+1}^m \Theta(d_{ij}-\lambda_i) } \ , 
 \label{eqn:eq2} 
\end{equation}
where $\Theta(x)$ is the Heaviside function.
Multiple-scattering contributions are included by summing all pairwise contributions (or double collisions) 
towards additional ionization.

We can derive the minimum and maximum possible values for $m_{\rm pix}$:
the minimum value is obtained in the PCM limit (the Heaviside function results in zero for all pairs):
\begin{equation}
m_{\rm pix}^{\min} = 1 \ ,
\end{equation}
and
\begin{equation}
m_{\rm pix}^{\max} = \frac{m+1}{2} 
\end{equation}
in the limit that the Heaviside function equals to one for all pairs.

The xPCM method allows for the following: along the classical trajectory the number of scattering events may increase.
For the first encountered atom we obtain the PCM-based primary scattering event; then, at the location of the second
encountered atom we have an additional, i.e., double scattering event with weight $1/m$, at the third encountered atom
a primary event plus a possible double scattering connected with the first atom, and a double-scattering event from
the second atom, etc..

\section{Results}
\label{sec:expt}

The atomic collision cross sections were calculated with the previously described TC-BGM program~Ref.~\cite{tcbgm}.
For proton projectiles these cross sections are shown in Fig.~1 of Ref.~\cite{hjl25a,hjl25b}, and for higher-charged projectiles 
in Fig.~1 of Ref.~\cite{hjl20a}. In this work it was shown how the PCM results obtained for projectile charges $Q=1,2,3$
can be scaled to higher charge values. TC-BGM calculations for projectile charge $Q=6$ were reported for
$\rm H_2O$ in Ref.~\cite{Pausz_2014}, and they are deemed reliable for impact energies greater than 40 keV/amu.

\subsection{Proton collisions}
\label{subsec:expt1}

\begin{figure}[h!]
  \centering
  \begin{minipage}[t]{0.35\textwidth}
   \centering
    \includegraphics[width=\textwidth]{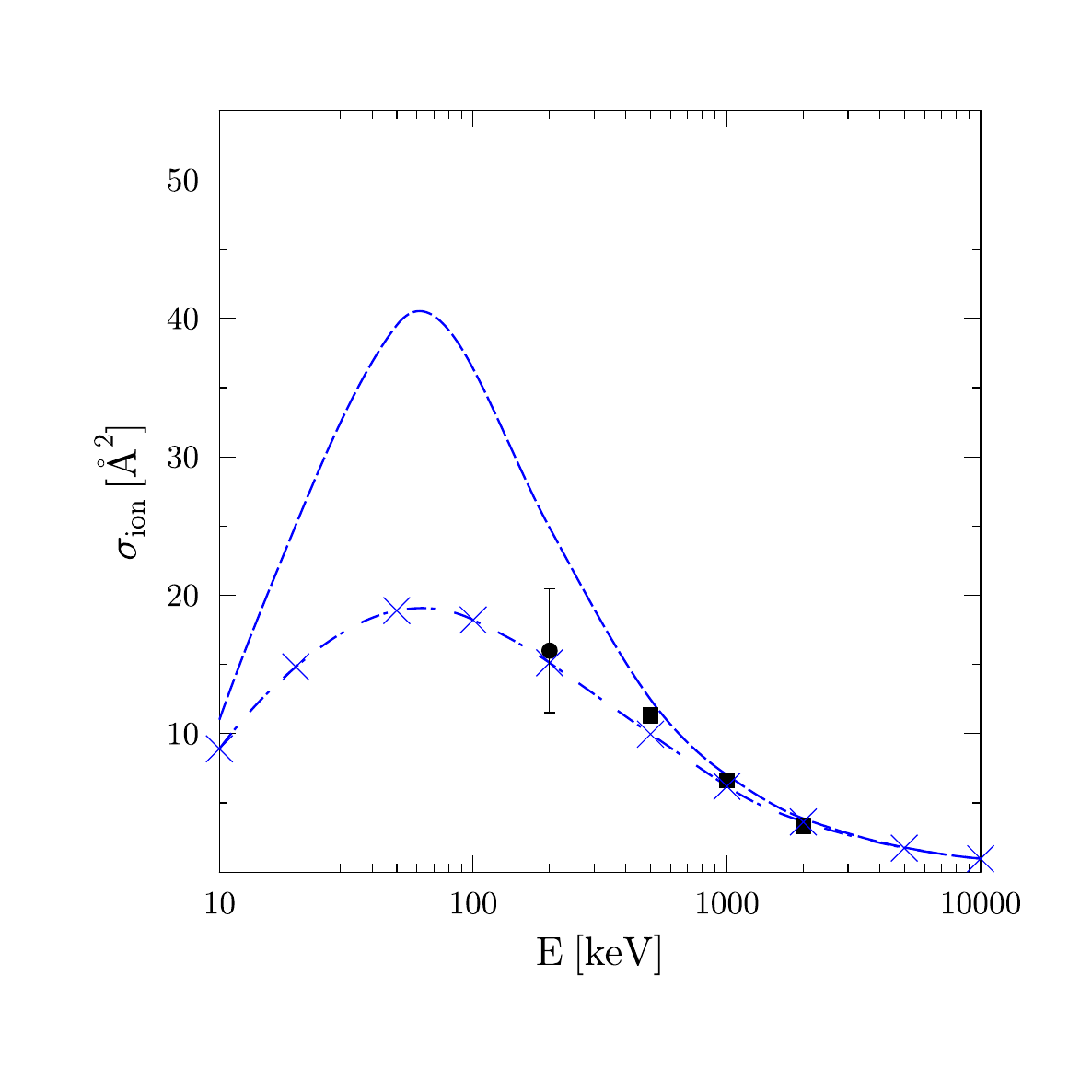}
  \end{minipage}
  \hskip -1 true cm 
  \begin{minipage}[t]{0.35\textwidth}
   \centering
    \includegraphics[width=\textwidth]{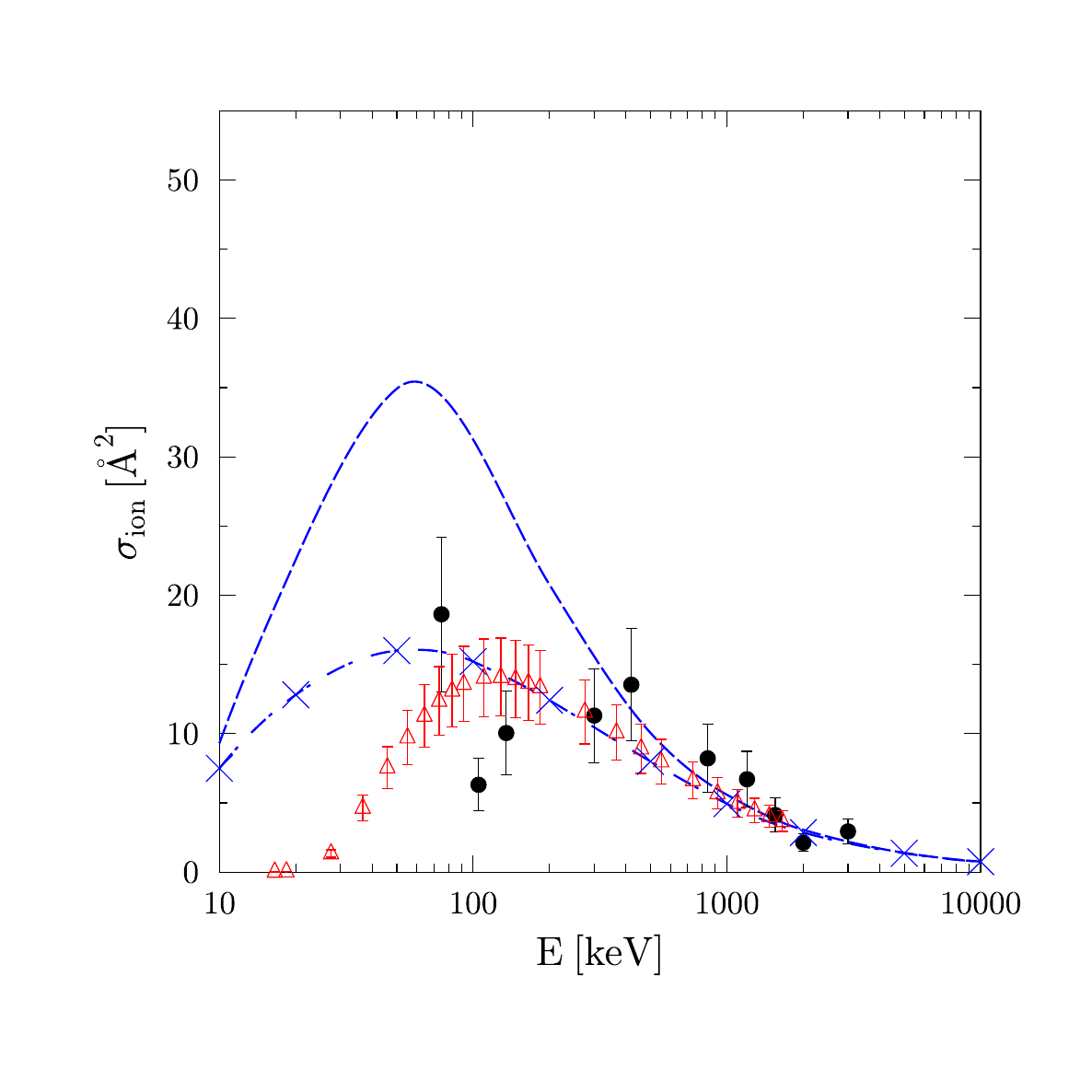}
   \end{minipage}
  \hskip -1 true cm 
   \begin{minipage}[t]{0.35\textwidth}
   \centering
    \includegraphics[width=\textwidth]{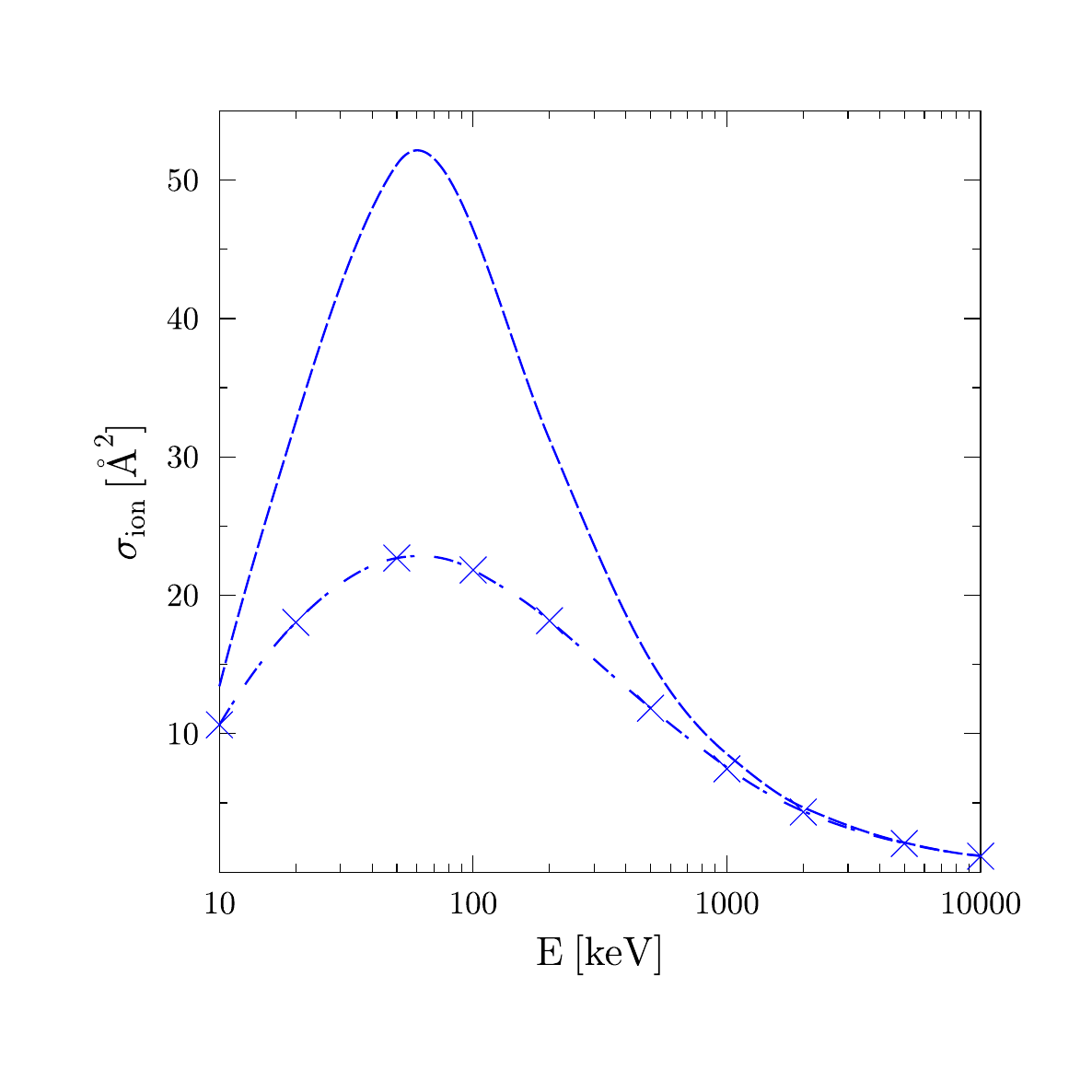}
  \end{minipage}
   \caption{Electron emission (net ionization) in proton collisions with uracil ($\rm C_4 H_4 N_2 O_2 $) (left panel), 
   tetrahydrofuran ($\rm C_4 H_8 O$) (middle panel),
   and valine ($\rm C_5H_{11} N O_2$) (right panel). Dashed lines: AR, dash-dotted lines: PCM, crosses: xPCM. 
   The experimental data in the left panel are from Ref.~\cite{Champion_2013} (solid squares) and Ref.~\cite{Chowdhury_2022} (solid circle).
   In the middle panel the proton-impact data are from Ref.~\cite{Rudek_2016} (solid circles), while the triangles 
   represent electron scattering data for equivalent impact velocity from Ref.~\cite{Bug_2017}.}
   \label{fig:emissionp+X} 
\end{figure}

In Fig.~\ref{fig:emissionp+X} results from xPCM are compared with PCM and AR for electron production (also referred to as net ionization) 
in proton collisions with three molecules. Obviously  xPCM agrees perfectly with PCM  at all collision energies.
Comparison with other theories, such as CDW-EIS and CTMC (among others) can be found in Fig.~5 of Ref.~\cite{hjl16}. 
For tetrahydrofuran (THF, $\rm C_4 H_8 O$) (middle panel) the comparison with experiment and the CNDO-Born calculation of Ref.~\cite{Rudek_2016}  
was previously shown in Fig.~4 of Ref.~\cite{hjl19}. 
The comparison with experimental electron data is relevant, 
since the experimental proton data at intermediate energies appear to be affected by large uncertainties.

The right panel of Fig.~\ref{fig:emissionp+X}  shows proton-valine ($\rm C_5 H_{11} N O_2$) electron production. In this case the number
of hydrogen atoms is larger and the total cross sections as obtained with all three IAM methods are larger than for uracil and THF.
Again, there is no significant difference between PCM and xPCM cross sections.

To summarize this section we can make the following observations: for proton collisions with molecules of linear extent in the
range of  3-4 \AA, and low atom number density, e.g., in the range $n <$ 0.1 \AA$^{-3}$, the probability for double-scattering
is very low for electron production, and this leads to the close agreement of PCM and xPCM results.
The situation would be quite different for single capture where in the energy range 10-50 keV/amu the cross sections
become large, and the corresponding mean free paths rather small.

\begin{figure}[h!]
 \centering
  \vspace{-1.3cm}
  \includegraphics[width=10cm]{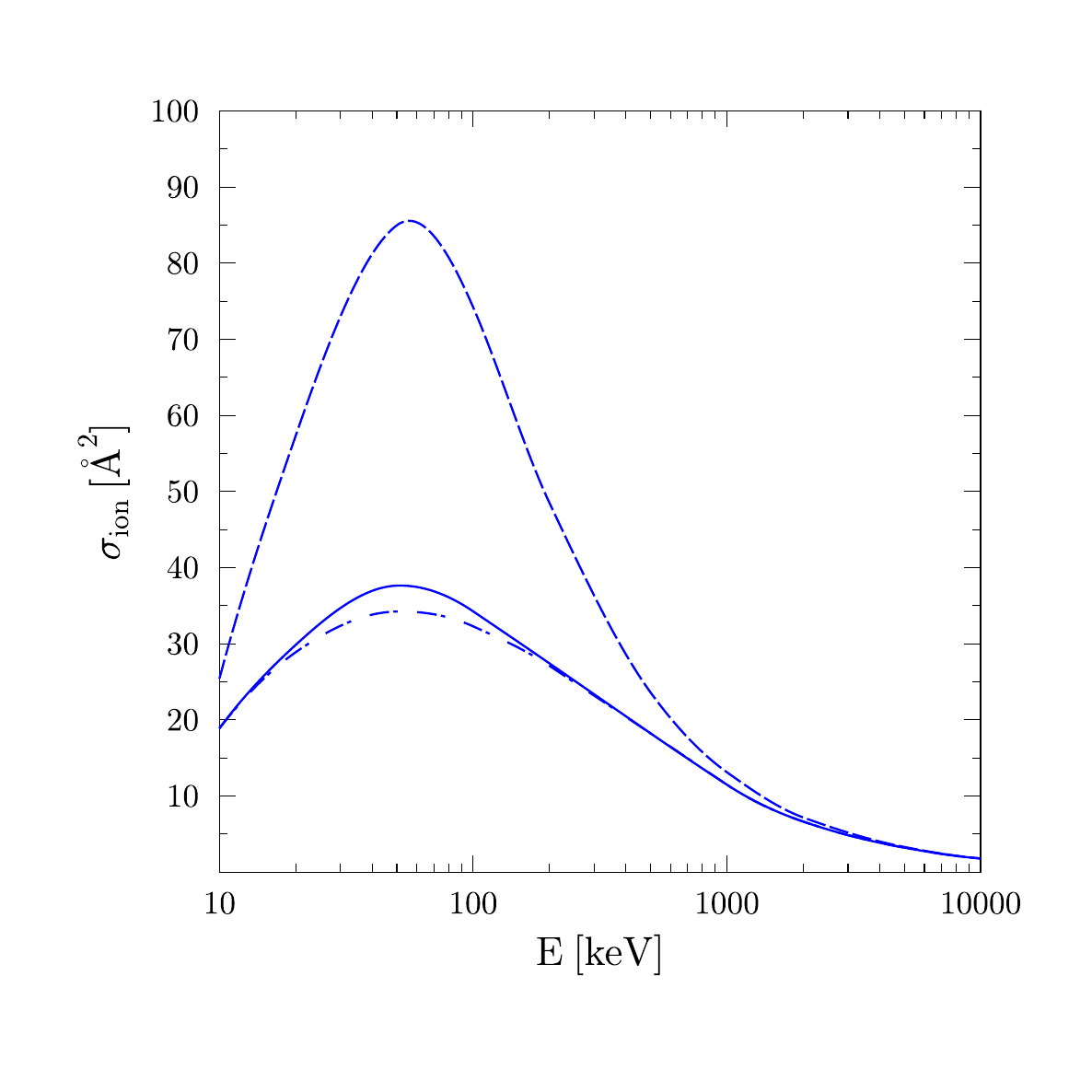} 
  \vspace{-1.5 cm}
 \caption{Electron production cross section for proton-anthracene ($\rm C_{14} H_{10}$) collisions from IAM calculations: dashed line: AR, dash-dotted line:
 PCM, solid line: xPCM.}
  \label{fig:p+anthracene} 
\end{figure}

Scattering from molecules with larger extent leads to sizeable double-scattering probabilities, particularly for atoms
which are further separated within the molecule. If one has a molecule with increased atomic number density for a
given species, e.g., carbon atoms in anthracene ($\rm C_{14} H_{10}$), then the mean free path is reduced sufficiently such that the 
PCM and xPCM results differ in the vicinity of the Bragg peak. Results for proton-anthracene collisions are shown in 
Fig.~\ref{fig:p+anthracene}. Net electron production as calculated with inclusion of double-scattering as per xPCM rises 
visibly above the PCM cross section.

\subsection{Highly charged ion collisions}
\label{subsec:expt2}

\begin{figure}[h!]
 \centering
  \vspace{-1.3cm}
  \includegraphics[width=10cm]{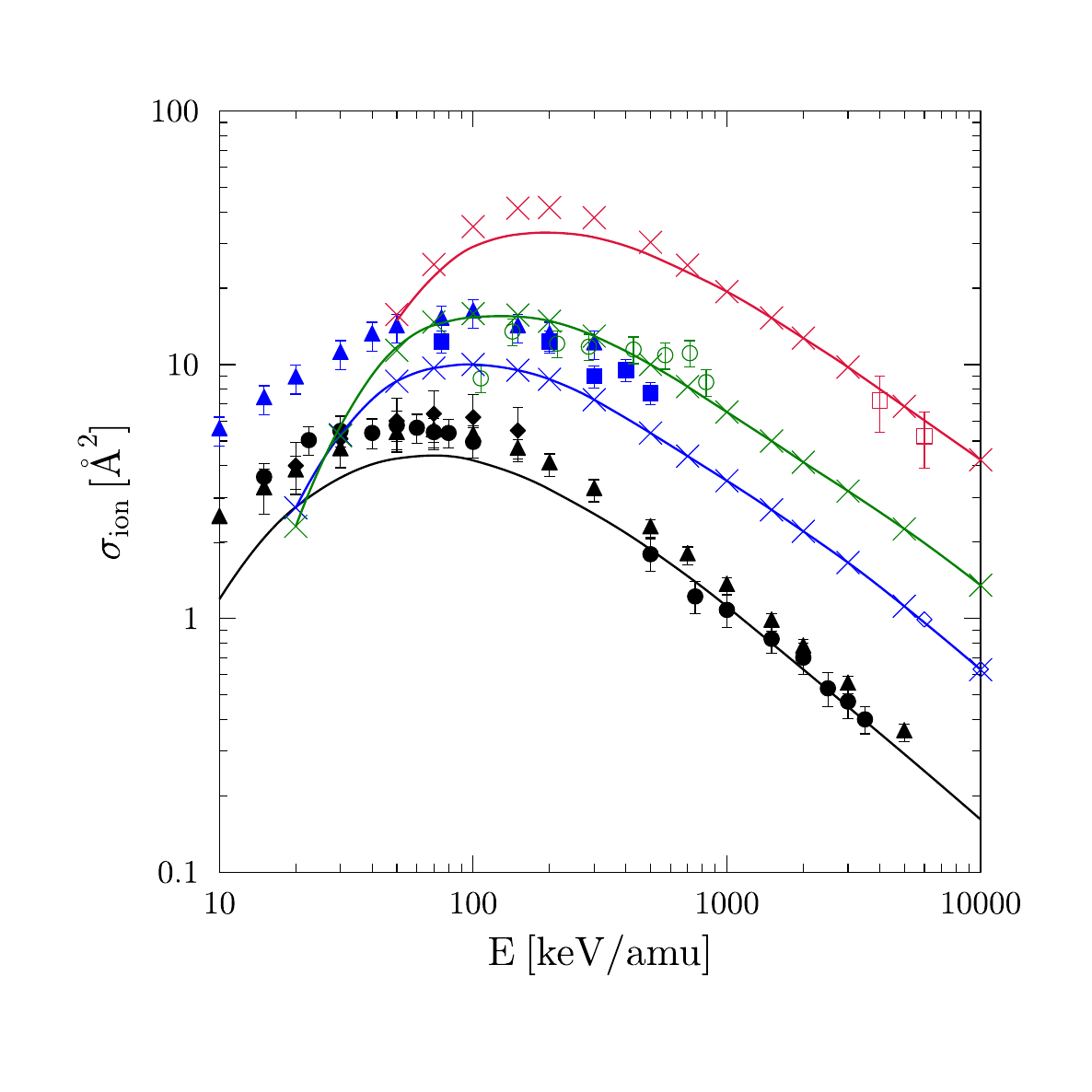} 
  \vspace{-1.5 cm}
 \caption{Electron emission in collisions of bare ions with projectile charges $Q=1,2,3,6$ from water molecules ($\rm H_2 O$).
 Solid lines: PCM, crosses: xPCM. Experimental data for $Q=1$ (black) diamonds: Ref.~\cite{Bolorizadeh86},
 squares: Ref.~\cite{Toburen_80}, triangles: Ref.~\cite{Rudd85c}, circles: Ref.~\cite{Luna07};  
 for $Q=2$: (blue) squares: Ref.~\cite{Toburen_80}, triangles: Ref.~\cite{Rudd85c}; for $Q=3$:
 (green) open circles: Ref.~\cite{Luna16}; for $Q=6$: (red) squares: Ref:~\cite{DALCAPPELLO2009781}.}
  \label{fig:H2O-ion} 
\end{figure}

For highly charged ion collisions we begin the discussion with a small molecule, namely water, for which we expect
double scattering to play a minor role. Fig.~\ref{fig:H2O-ion} shows the PCM results given in Fig.~9 of Ref.~\cite{hjl20a}.
It is apparent that for projectile charges $Q=1,2,3$ no significant enhancement of the PCM electron production cross section is obtained
from double scattering as modelled in xPCM, but that an enhancement occurs near the Bragg peak for bare carbon
projectiles ($Q=6$). Note that this enhancement increases the electron production by about $25-30 \%$ at the maximum.
The reason for this double-scattering effect
is the decrease in the mean free path for bare carbon ion impact such that it becomes less than the bond length
between the oxygen and hydrogen atoms.
Away from the maximum, however, the two models predict the same cross section values. This happens when the
cross section values are at the level of about 20~\AA${}^2$, or lower. 

\begin{figure}[h!]
  \centering
  \begin{minipage}[t]{0.5\textwidth}
   \centering
   \includegraphics[width=\textwidth]{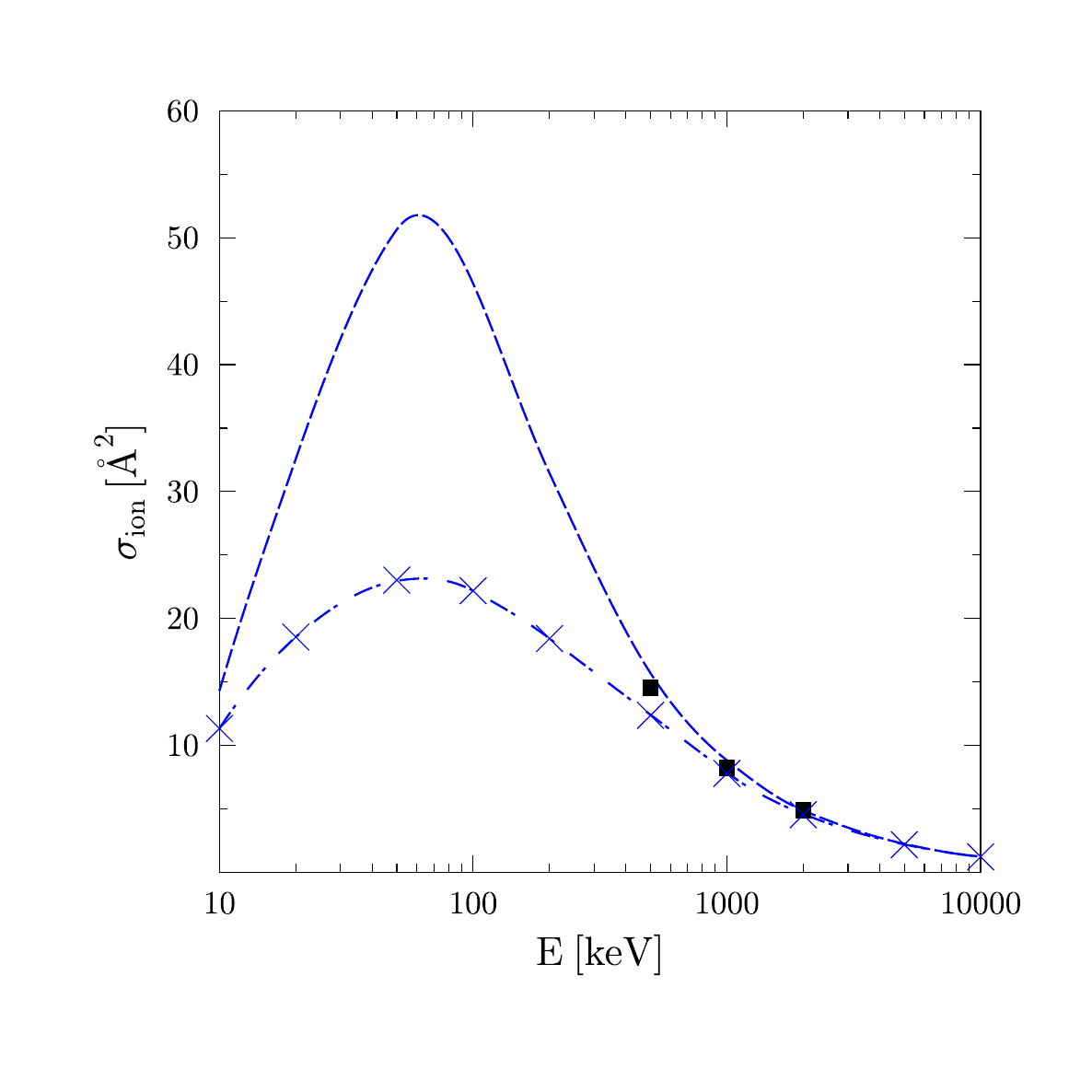}
  \end{minipage}
  \hskip -1 true cm 
   \begin{minipage}[t]{0.5\textwidth}
   \centering
   \includegraphics[width=\textwidth]{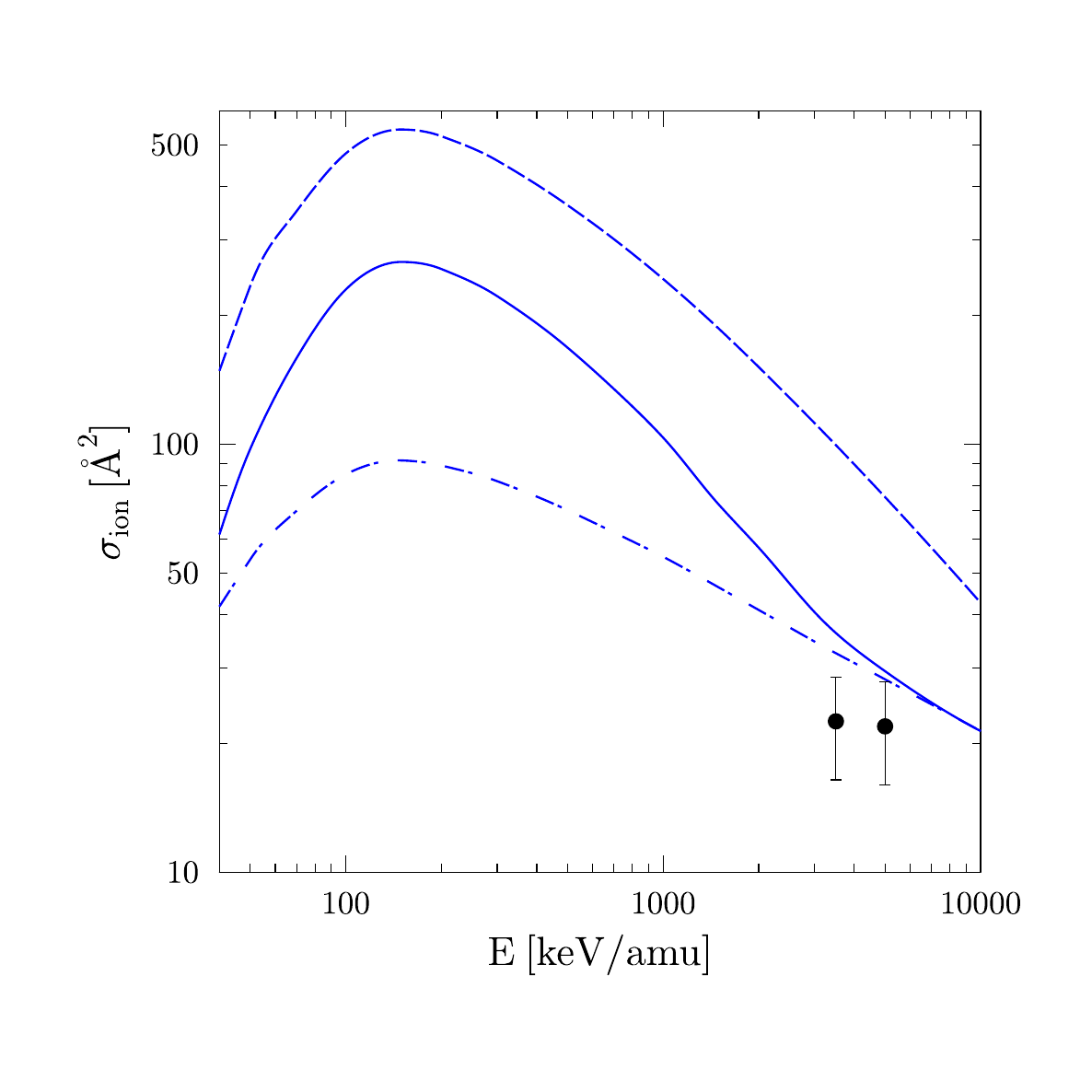}
  \end{minipage}
   \caption{Electron emission in scattering from adenine ($\rm C_5 H_5 N_5$). Left panel: proton impact, right panel: bare carbon ion impact.
   Dashed lines: AR, Dash-dotted lines: PCM, Crosses x-PCM. Experimental data are from Ref.~\cite{Iriki_11} for protons,
   and Ref.~\cite{Bhattacharjee_2020} for bare carbon ions.
    }
   \label{fig:emissionp+Adenin} 
\end{figure}

We now move on to the bases adenine ($\rm C_5 H_5 N_5$) and uracil ($\rm C_4 H_4 N_2 O_2$). 
They both have a compact structure with a maximum linear extent of about 6.6~\AA~for
adenine and 5.2~\AA~for uracil. For adenine comparison with experimental data is shown in Fig.~\ref{fig:emissionp+Adenin}.
For proton impact (left panel) we observe that PCM and xPCM produce practically identical results over the entire range
of collision energies. The experimental data of  Ref.~\cite{Iriki_11} are reproduced for the two highest energies, but the
data point at 500 keV falls closer to the AR result. The experimental data were obtained by integrating
doubly-differential cross sections, and have very small statistical uncertainties. In the right panel the data for
$\rm C^{6+}$ projectiles are presented on a double-logarithmic scale, and are compared with
the recent measurements of Ref.~\cite{Bhattacharjee_2020}. 

We observe a dramatic increase in electron emission due to the modelling of multiple scattering in xPCM,
and note that the shape of the curve approaches that of AR in a wide Bragg peak.
Comparison with other theoretical calculations was provided in Fig.~3 of Ref.~\cite{hjl16}.

\begin{figure}[h!]
  \centering
  \begin{minipage}[t]{0.5\textwidth}
   \centering
   \includegraphics[width=\textwidth]{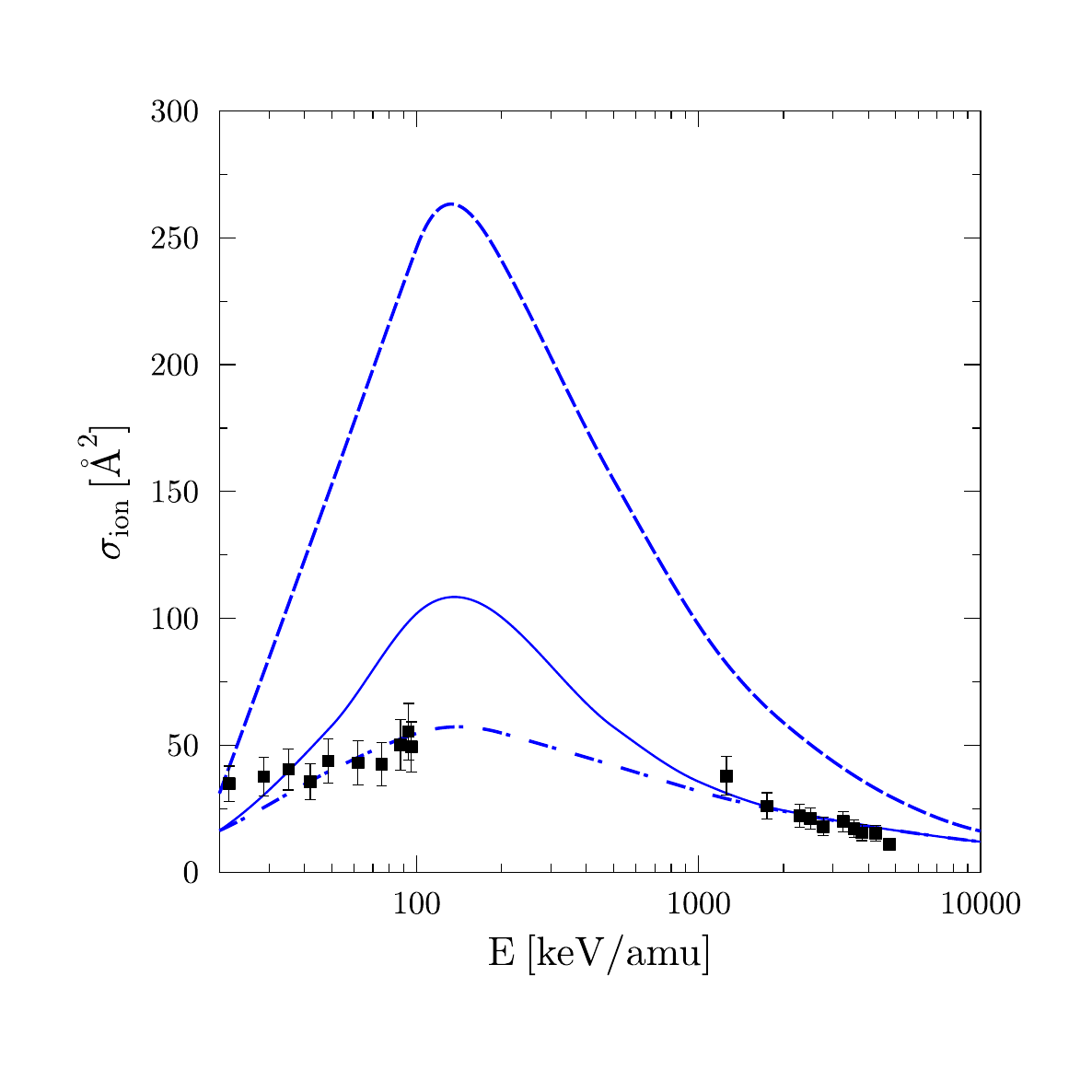}
  \end{minipage}
  \hskip -1 true cm 
   \begin{minipage}[t]{0.5\textwidth}
   \centering
    \includegraphics[width=\textwidth]{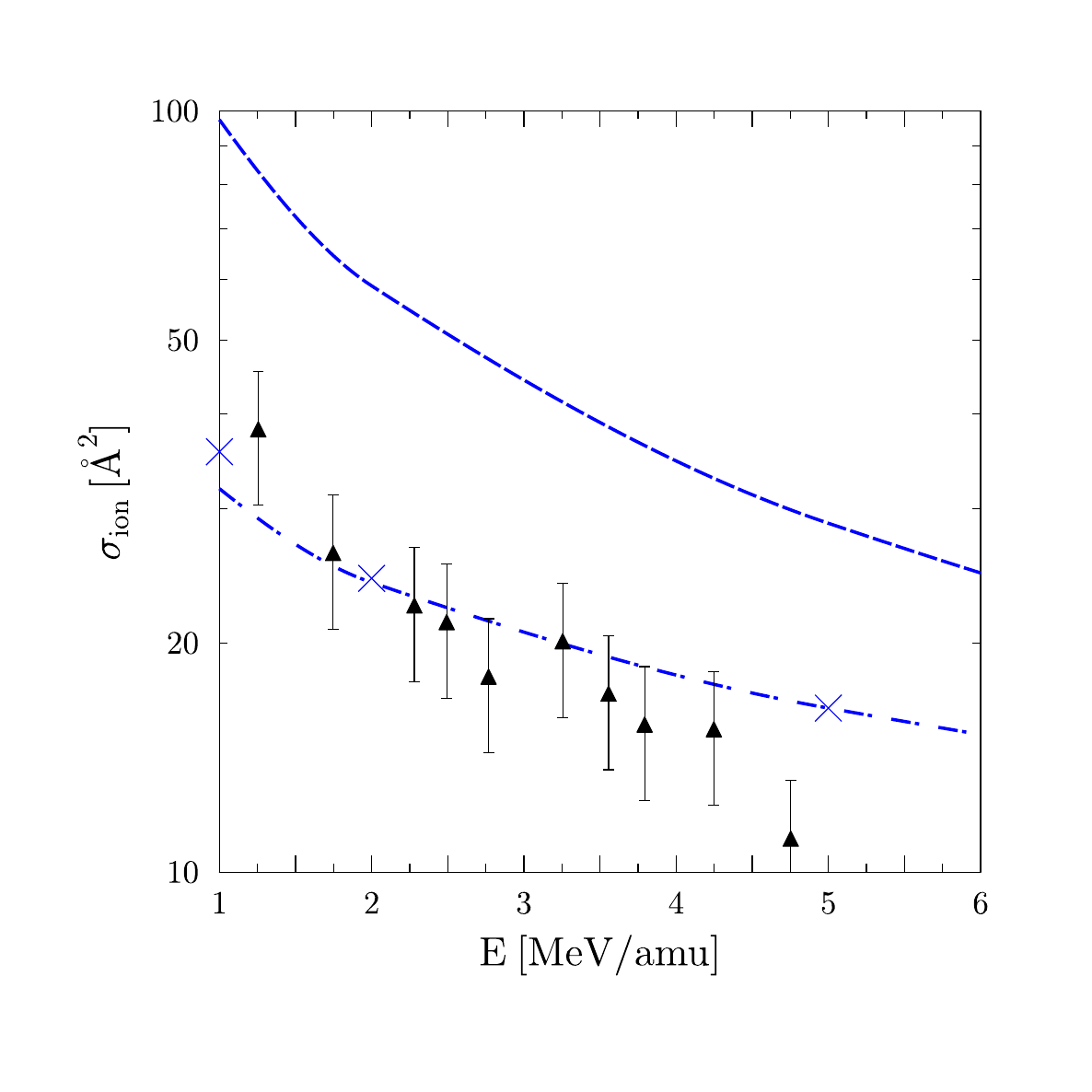}
  \end{minipage}
   \caption{Electron emission in C$^{4+}$ collisions with uracil ($\rm C_4 H_4 N_2 O_2$).  Dashed lines: AR, dash-dotted lines: PCM. Results from xPCM
   are shown as a solid line in the left panel, and as crosses in the right panel. Experimental data: left panel (squares): 
   Ref.~\cite{PhysRevA.85.032711};
   right panel (triangles): the high-$E$ data from Ref.~\cite{PhysRevA.85.032711} shown in the left panel.}
   \label{fig:emissionp4+Uracil} 
\end{figure}

For uracil targets there are more data, and particularly for a wider energy range. We first show data for $\rm C^{4+}$
projectiles which are treated theoretically by using a screened projectile potential, and which are shown in Fig.~\ref{fig:emissionp4+Uracil}.
The left panel shows them on a linear vs logarithmic scale for $\sigma_{\rm ion}(E)$, and demonstrates a dramatic increase in the Bragg peak region
of double-scattering as described by xPCM. At high energies where xPCM and PCM results coalesce the agreement
with experiment is very good. The right panel shows this region with logarithmic vs linear axis scaling for 
$\sigma_{\rm ion}(E)$ vs $E$. Apart from the highest experimental energy data point the agreement is within
one standard deviation of the experiment. The lowest-$E$ data point has the potential of supporting xPCM over PCM,
but more data at slightly lower energies would be required for confirmation.

At the low-$E$ end of the data (left panel), between 20 and 100 keV/amu the comparison of the present data with
experiment is inconclusive, but it seems that the approach of the Bragg peak region is described better by PCM than
by xPCM. In the region of the lowest-$E$ data points where PCM and xPCM agree both models have a tendency
of being lower than experiment (cf. Fig.~\ref{fig:H2O-ion} for water molecule targets).

\begin{figure}[h!]
  \centering
   \begin{minipage}[t]{0.5\textwidth}
   \centering
    \includegraphics[width=\textwidth]{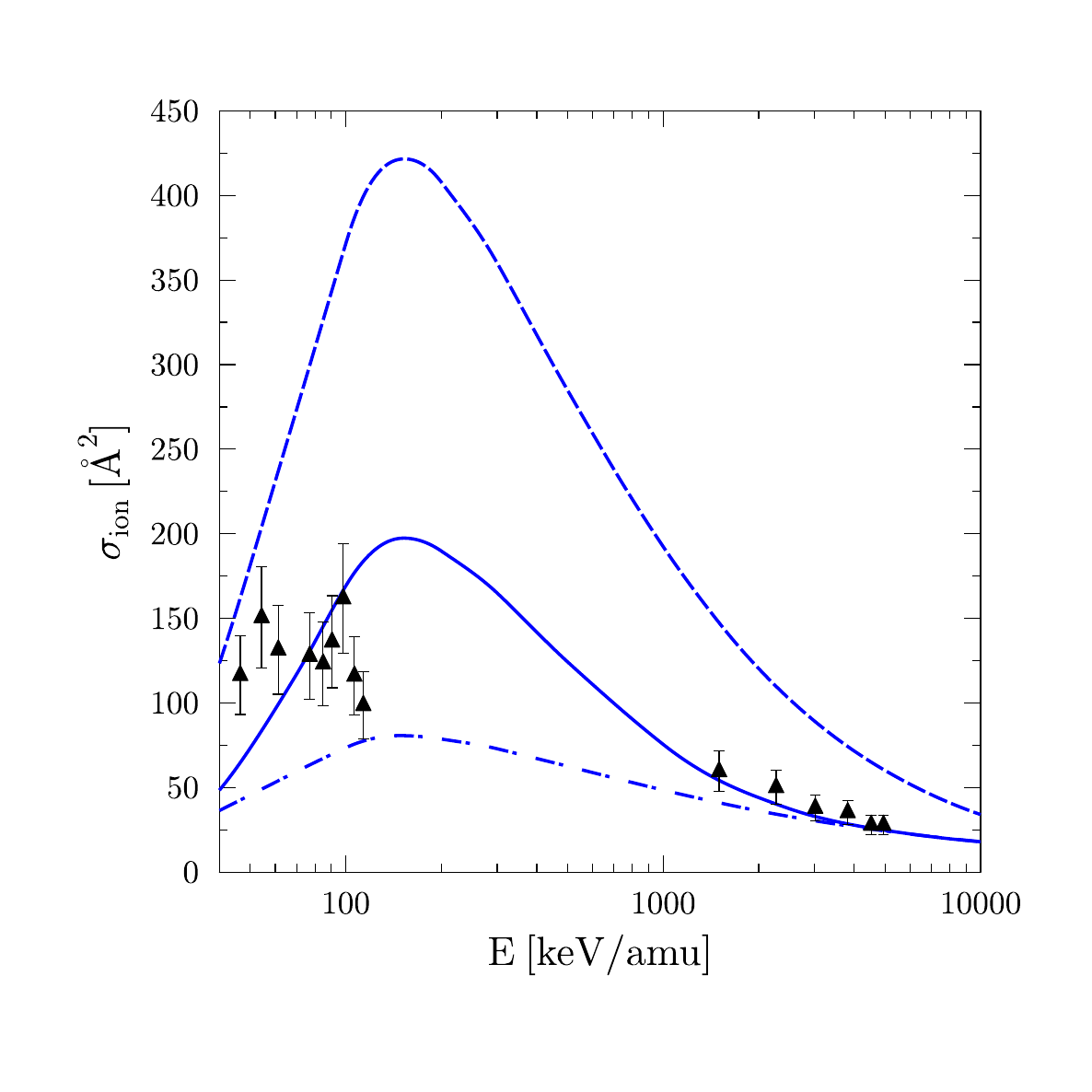}
  \end{minipage}
  \hskip -1 true cm 
   \begin{minipage}[t]{0.5\textwidth}
   \centering
    \includegraphics[width=\textwidth]{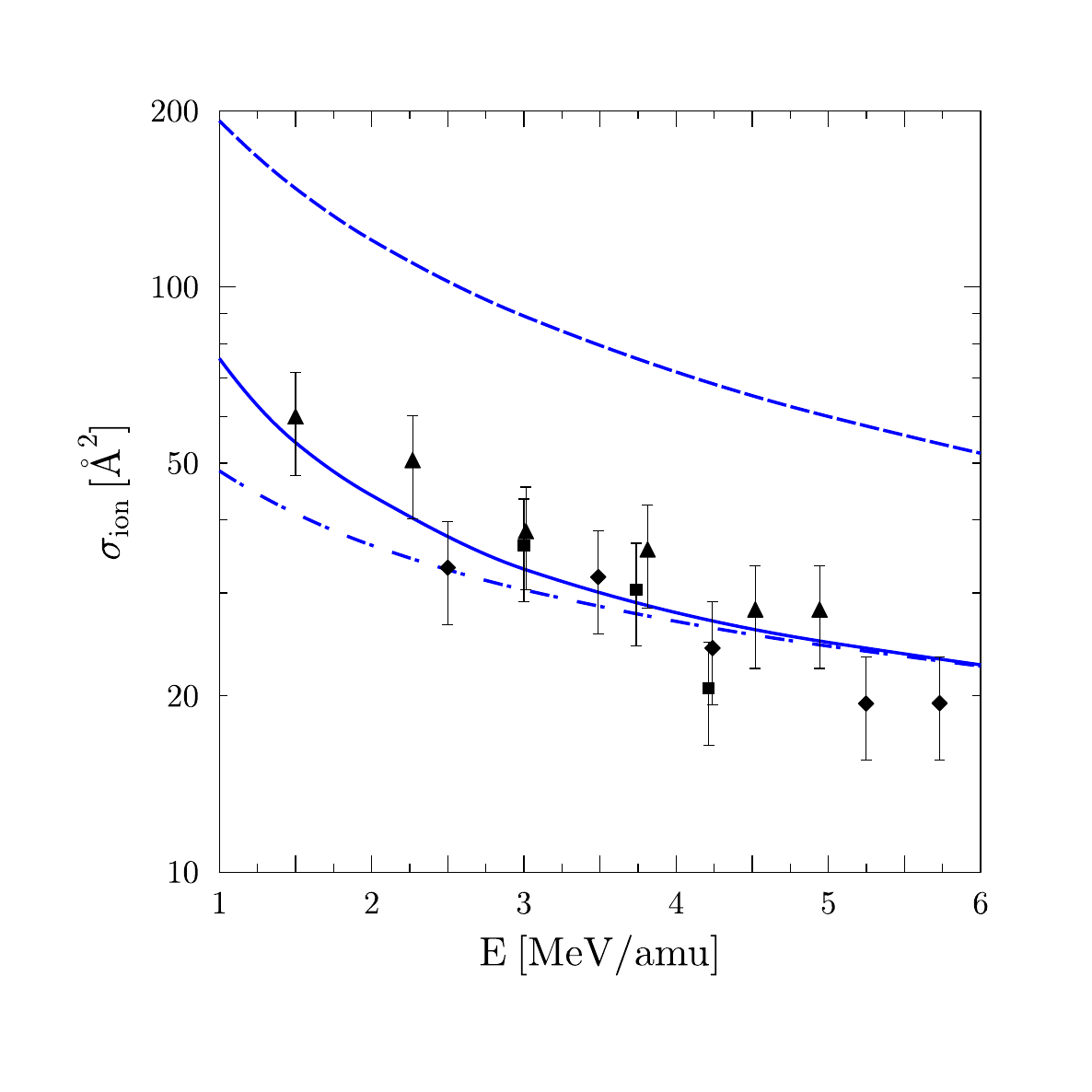}
  \end{minipage}
   \caption{Electron emission in C$^{6+}$ collisions with uracil ($\rm C_4 H_4 N_2 O_2$).  Dashed lines: AR, dash-dotted lines: PCM, solid lines: xPCM. Experimental data
   from  Ref.~\cite{PhysRevA.85.032711}
   are for bare carbon (right panel, diamonds), for $\rm F^{6+}$(solid squares) and for $\rm O^{6+}$ (triangles).}
   \label{fig:emissionp6+Uracil} 
\end{figure}

For bare carbon ion impact results are shown in Fig.~\ref{fig:emissionp6+Uracil}.
Here we observe in the linear-logarithmic presentation of $\sigma_{\rm ion}(E)$
that xPCM falls into the correct range at low energies, but the scatter of the data is substantial.
At high energies the xPCM result is supported very well by experiment.

In the right panel $\sigma_{\rm ion}(E)$ is presented in log-linear fashion for the high-$E$ range for
which $Q=6$ projectile ions C, O, F are available from experiment. These experimental data
agree within their statistical error bars. It is interesting that these data with the lowest-$E$ points
for $\rm O^{6+}$ lend support to the xPCM over the PCM results. We hope that a more conclusive
answer will come in the future from experimental data in the Bragg peak region.

\begin{figure}[h!]
  \centering
  \begin{minipage}[t]{0.5\textwidth}
   \centering
    \includegraphics[width=\textwidth]{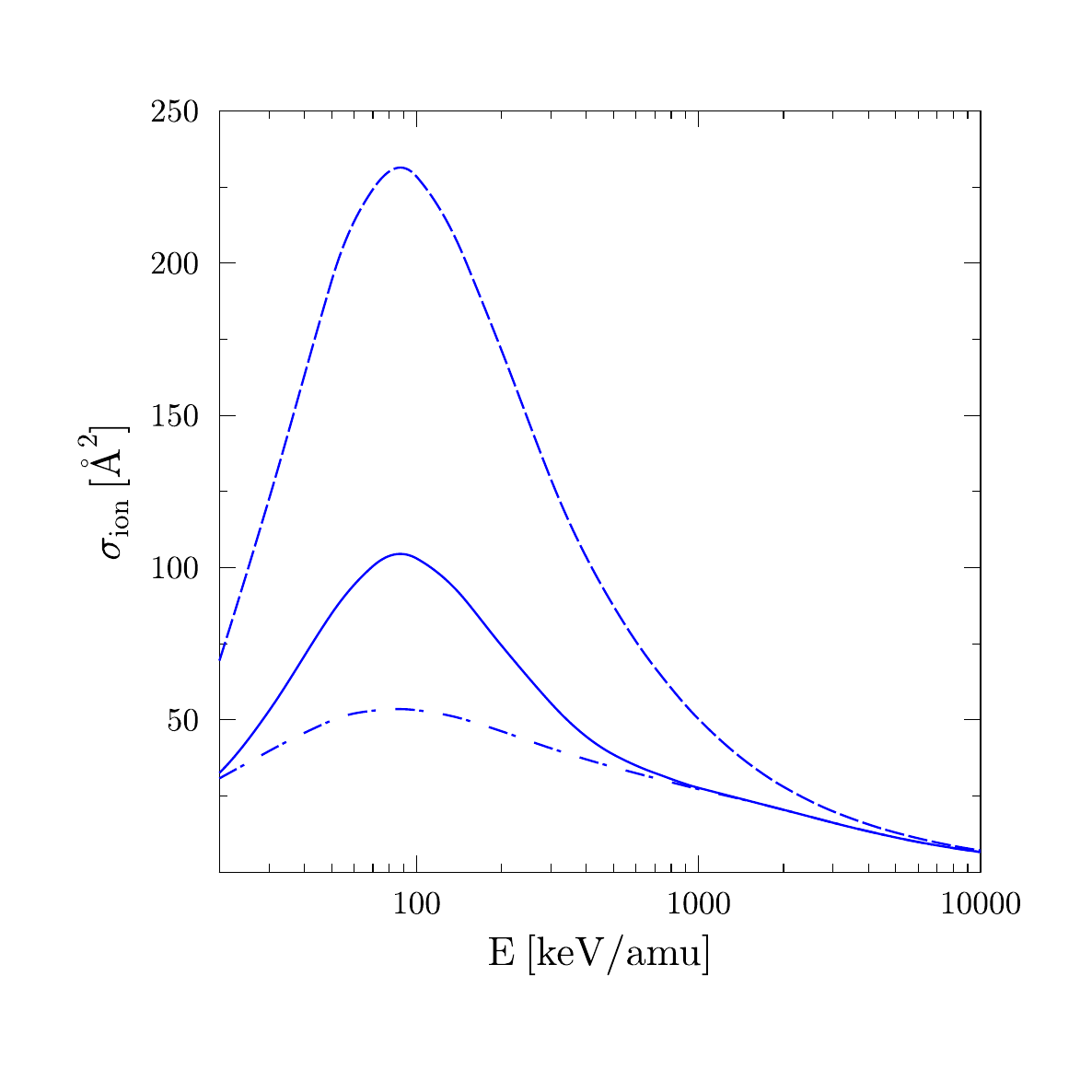}
  \end{minipage}
  \hskip -1 true cm 
   \begin{minipage}[t]{0.5\textwidth}
   \centering
    \includegraphics[width=\textwidth]{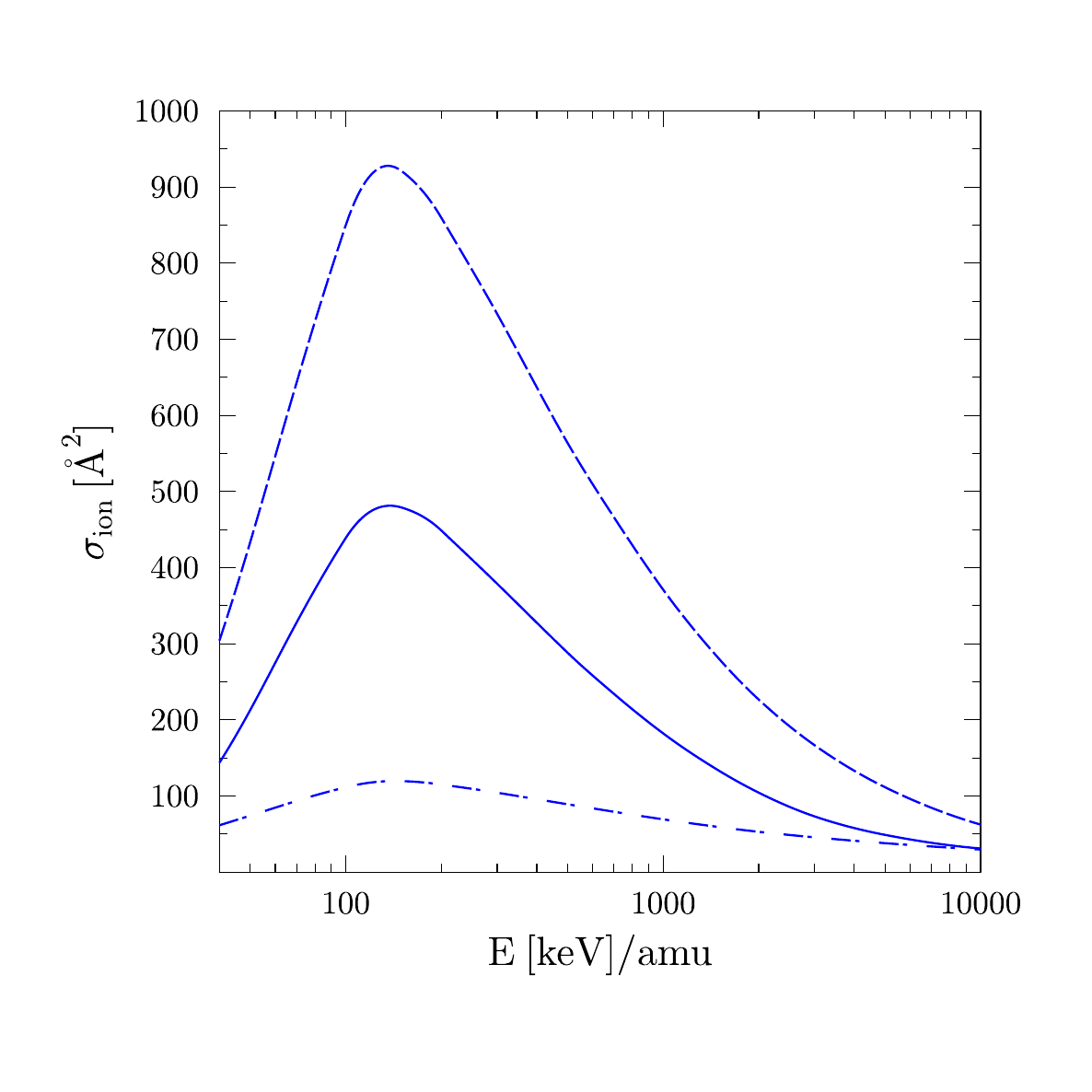}
  \end{minipage}
   \caption{Electron emission in collisions with anthracene ($\rm C_{14} H_{10}$) for $\rm He^{2+}$ projectiles in the left panel, and $\rm C^{6+}$ projectiles in the right panel.}
   \label{fig:emissionp+Anthracen} 
\end{figure}

We end this section with an example of an even larger molecule, namely anthracene ($\rm C_{14} H_{10}$) with linear extension
approaching 10~\AA~shown in Fig.~\ref{fig:emissionp+Anthracen}. As was shown in Fig.~\ref{fig:p+anthracene}
even proton collisions allow for double scattering in the vicinity of the Bragg peak for this molecule.
In the left panel of Fig.~\ref{fig:emissionp+Anthracen} we show that for $Q=2$ projectiles the enhancement 
of electron production due to the inclusion of double scattering results in a doubling of the cross section
at the Bragg peak when considering the xPCM vs the PCM result. For $Q=6$ (right panel) the enhancement 
leads to an increased electron emission which is about half way between the PCM and the AR result.
Our conclusion, therefore, is that experimental data in the range of 50-300 keV/amu for $\rm He^{2+}$ projectiles,
and an even wider range for ions with higher charge states are needed to confirm the multiple-scattering effect.

\section{Conclusions}
\label{sec:conclusions}

We have developed and applied a semiclassical multiple-scattering approach to IAM descriptions of ion-molecule collisions.
The implementation is carried out using the PCM technique by considering pairs of atoms within the molecule.
The current situation of experimental data is not yet sufficient to draw a definite conclusion, but the indication
from some comparisons made in this paper points to the existence of the effect. This is not only a challenge for PCM,
but the existence of such an effect in ionizing heavy-ion collisions can become a challenge for other independent-electron
methods which are currently in use for explaining experimental data, and in the modelling of therapy by ion impact
in radiation medicine.

\begin{acknowledgments}
We would like to thank the Center for Scientific Computing, University of Frankfurt for making their
High Performance Computing facilities available.
Financial support from the Natural Sciences and Engineering Research Council of Canada 
(Grants No. RGPIN-2023-05072 and No. RGPIN-2025-06277)
is gratefully acknowledged. \end{acknowledgments}

%

\bibliography{xPCM}

\end{document}